\documentclass[useAMS,usenatbib]{mn2e} \input{epsf}

\newif\ifAMStwofonts 

\def\lesssim{\mathrel{\hbox{\rlap{\hbox{\lower4pt\hbox{$\sim$}}}\hbox{$<$}}}}
\def\gtrsim{\mathrel{\hbox{\rlap{\hbox{\lower4pt\hbox{$\sim$}}}\hbox{$>$}}}}
\def\msun{${\rm M}_{\odot}$~}

\def\ll_lsun{$\log{L/\rm L_{\odot}}$~}
\def\masa_msun{$M/ \rm M_{\odot}$~}

\def\m_mstar{$M/M_{*}$~}

\def\aap{A\&A}
\def\apj{ApJ}
\def\apjl{ApJ}
\def\mnras{MNRAS}

\title[Formation of Heium White  Dwarf in Binary Evolution]  {The 
Formation  of  a  Helium  White Dwarf in Close Binary
System with Diffusion}

\author[O.   G.   Benvenuto, M.   A.   De   Vito]   
{
O.~G.
Benvenuto$^{1,2}$\thanks{Member   of   the   Carrera   del   Investigador
Cient\'{\i}fico, Comisi\'on  de Investigaciones  Cient\'{\i}ficas
de   la   Provincia   de   Buenos   Aires (CIC),   Argentina.    Emails:
obenvenuto@fcaglp.unlp.edu.ar; obenvenu@astro.puc.cl}
M.~A.  De Vito$^{1}$\thanks{Fellow   of   the   CIC. Email: adevito@fcaglp.unlp.edu.ar}
\\     
$^{1}$ Facultad de Ciencias Astron\'omicas y Geof\'{\i}sicas, Universidad Nacional de La
Plata, Paseo del Bosque S/N, (1900) La Plata, Argentina\\
$^{2}$ Departamento de Astronom\'{\i}a y Astrof\'{\i}sica, Pontificia
Universidad Cat\'olica, Vicu\~na Mackenna 4860, Casilla 306, Santiago, Chile}

\begin{document}

\date{December 30}

\pagerange{\pageref{firstpage}--\pageref{lastpage}} \pubyear{2003}

\maketitle \label{firstpage}

\begin{abstract} 

We study  the evolution  of a system  composed by a  1.4~\msun neutron
star and a  normal, solar composition star of 2~\msun  in orbit with a
period  of 1 day.   Calculations were  performed employing  the binary
hydro code presented in Benvenuto \& De Vito (2003) that handle
the mass  transfer rate in a  fully implicit way. Now  we included the
main standard  physical ingredients together  with diffusion processes
and  a proper  outer boundary  condition.  We  have assumed  fully 
non~-~conservative mass transfer episodes.

In order to study the interplay of mass loss episodes and diffusion we
considered evolutionary sequences with  and without diffusion in which all
Roche  lobe overflows (RLOFs)  produce mass transfer.   Another two
sequences  in which thermonuclearly-driven RLOFs are not allowed to  drive
mass transfer have been computed with and without diffusion. To our
notice, this study represents the first binary evolution calculations in
which diffusion is considered.

The system  produces a  helium white dwarf  of $\sim 0.21$~\msun  in an
orbit with a  period of  $\sim 4.3$~days for the  four cases.  We find 
that mass transfer episodes induced by hydrogen thermonuclear flashes
drive a  tiny  amount of  mass  transfer.   As  diffusion produces 
stronger flashes, the amount of hydrogen  - rich matter transferred is
slightly higher than in models without diffusion. 

We find that diffusion is
the main agent  in determining the evolutionary timescale  of low mass
white dwarfs even in presence of mass transfer episodes.

\end{abstract}


\section{Introduction} \label{sec:intro}

At present,  it is a well  established fact that low  mass white dwarf
(WD)  stars should  be formed  during  the evolution  in close  binary
systems  (CBSs).  These  objects are  expected to  have a  helium rich
interior  simply because  they have  a  mass below  the threshold  for
helium  ignition  of about  $0.45~M_\odot$.  If  they  were formed  as
consequence  of single  star  evolution,  we would  have  to wait  for
timescales far  in excess of the  present age of the  Universe to find
some of them.

Formation  of  helium WDs  in  CBSs  was  first investigated long ago by
Kippenhahn, Kohl  \& Weigert (1967) and Kippenhahn,  Thomas \& Weigert
(1968). They found that these  objects are formed during the evolution
of low mass CBSs and that the cooling evolution is suddenly stopped by
thermonuclear flashes  that are able to  swell the star  up to produce
further Roche lobe overflows (RLOFs). 

Since sometime ago, low mass WDs have been discovered as companions to
millisecond pulsars  (MSP). This fact sparkled interest  in helium WDs in
order to  investigate the deep physical links  between both members of  a 
given  pair.    In  particular,  it  represents  an  attractive
possibility to infer characteristics of the neutron star behaving as a
MSP  by  studying the  WD  in detail.  Studies  devoted  to helium  WD
properties are those  of Alberts, et al. (1996);  Althaus \& Benvenuto
(1997); Benvenuto  \&  Althaus  (1998);  Hansen  \&  Phinney  (1998a);
Driebe, et al. (1998); Driebe, et al. (1999); Sch{\"o}nberner, Driebe,
\&   Bl{\"o}cker  (2000);  Althaus   \&  Benvenuto   (2000);  Althaus,
Serenelli, \& Benvenuto (2001abc); Serenelli, et al. (2001); Rohrmann, et 
al. (2002);  Serenelli, et  al.  (2002). Sarna,  Ergma, \&  Ger{\v
s}kevit{\v s}-Antipova  (2000) considered the problem in  the frame of
detailed binary evolution  calculations. More recently, Podsiadlowski, 
Rappaport, \& Pfahl (2002) have   also  computed   the  evolution   of   
some  CBS configurations that give rise to the formation of helium WDs.
Also, Nelson \& Rappaport (2003) have explored in detail the evolutionary
scenarios of binary systems with initial periods shorter than the
bifurcation one leading to the formation of ultra-compact binaries with
periods shorter than an hour. On the opposite, in this paper we shall deal
with a system with an initial period larger than the bifurcation one
leading to wider binaries.

Remarkably, the first WD found as companion of a MSP in globular clusters has been detected by Edmonds et
al. (2001). Among recent observations  of low mass WD companion  to MSPs we should
quote  those by  van Kerkwijk  et al.  (2000) who  have detect  the WD
companion  of  the  binary  MSP  PSR  B1855+09,  whose  mass  is  know
accurately  from  measurements of  the  Shapiro  delay  of the  pulsar
signal, $M_{WD}= 0.258^{+0.028}_{-0.016}$~\msun. The orbital period of
this binary  MSP is 12.3 days.  More recently, Bassa,  van Kerkwijk \&
Kulkarni (2003) have  found a faint bluish counterpart  for the binary
MSP  PSR J02018+4232.   The spectra  confirm that  the companion  is a
helium WD and, in spite  that observations are of insufficient quality
to  put  a strong  constrain  on the  surface  gravity,  the best  fit
indicates  a  low  $\log{g}$   value  and  hence  low  mass  ($\approx
0.2$~\msun).  On the other  hand, independently, Ferraro et al. (2003)
and Bassa et  al. (2003) have identified the  optical binary companion
to  the MSP  PSR  J1911-5958A, located  in  the halo  of the  galactic
globular cluster  NGC 6752. This object  turned out to be  a blue star
whose position  in the color-magnitude diagram is  consistent with the
cooling  sequence  of   a  low-mass  ($\approx~0.17-0.20$~\msun),  low
metalicity  helium WD  at the  cluster distance.  This is  the second
helium WD with a mass in this range that has been found to orbit a MSP
in a galactic globular clusters.  Also, Sigurdsson, et al. (2003) have
detected two companions for the  pulsar B 1620-26, one of stellar mass
and  one of planetary  mass. The  color and  magnitude of  the stellar
companion    indicate   a   WD    of   $0.34\pm0.04$~\msun    of   age
$4.8~\times~10^{8}$~y. For previous detections of this kind of objects we refer the reader to
the paper by Hansen \& Phinney (1998b).

From a  theoretical point of view,  it was soon realized  that the key
ingredient of WD models is the hydrogen mass fraction in the star.
Consequently, this called for a detailed treatment of the outer layers
of the  star.  Iben \&  MacDonald (1985) demonstrated the  relevance of
diffusion in the  evolution of intermediate mass CO  WDs while Iben \&
Tutukov (1986) found it to be also important in low mass WDs.

More recently, Althaus, et al.  (2001abc) revisited the problem of the
formation of helium  WDs. In doing so, they  mimicked binary evolution
by abstracting mass  to a 1~\msun object on the RGB.  The main goal of
these papers was to investigate in detail the role of diffusion during
the evolution as a pre-WD object. They allowed gravitational settling,
chemical  and thermal  diffusion  to operate.  However,  they did  not
considered  the  possibility  of   any  mass  transfer  episode  after
detachment from the  RGB.  Perhaps the main result  of Althaus, et al.
(2001abc) was the finding that  for models with diffusion there exists
a  threshold mass  value  $M_{th}$ above  which  the object  undergoes
several  thermonuclear  flashes  in  which  a large  fraction  of  the
hydrogen present in  the star is burnt out.  Consequently, as the star
enters on the  final cooling track it evolves  fast, reaching very low
luminosities  on   a  timescale  comparable   with  the  age   of  the
Universe. Quite contrarily,  in models without diffusion, evolutionary
timescales  are much  longer, making  it difficult  to  reconcile with
observations.  For WDs belonging  to CBSs  in companion
with MSP,  WD ages should be  comparable to the  characteristic age of
pulsars  $\tau_{PSR}= P/2\dot{P}$  (for  a
pulsar  of period  $P$ with  period derivative  $\dot{P}$ that  had an
initial  period  $P_0$  such  that  $P_0 << P$  and  braking  index
$n=3$). This should  be so, because it is  generally accepted that the
MSP is recycled  by accretion from its normal  companion.  However, it
was found that  the WD was much dimmer than predicted  by 
models without diffusion,  which  should  be  interpreted  as consequence  of  a  faster
evolution.  This  has been the  case of the
companion of PSR B1855+09.

For  objects with  masses  below 
$M_{th}$ no thermonuclear  flash occurs and the star  does not suffer
from another  RLOF.  Consequently, it retains a  thick hydrogen layer,
able to support nuclear burning, forcing the WD to remain bright for a
very long time. This is the case of the companion of PSR J1012+5307.

It  is the aim  of the  present paper  to revisit  the problem  of the
formation  and evolution  of helium  WDs  in CBSs  by performing  full
binary computations considering diffusion  starting with models on the
main sequence all  the way down to stages of  evolution of the remnant as
a very cool  WD. To our notice this is the  first time such a study is
carried out. In this way we largely generalize the previous studies from 
our group  on this  topic.  In doing  so, we  have preferred  to
concentrate  on  a  particular  binary system,  deferring  a  detailed
exploration  of the  huge  parameter space  (masses, orbital  periods,
chemical compositions, etc.)  to  future publications. To be specific, we 
have chosen  to study  a  CBS composed  by a  2~\msun normal  star
together with a neutron star with a ``canonical'' mass of 1.4~\msun on an 
initial orbit  of 1  day of  period. We assumed solar chemical 
composition with $Z=0.02$   for    which $M_{th}~\approx~0.19$~\msun
(Althaus et al. 2001b).

In order to  explore the role and interplay of  mass loss episodes and
diffusion  we  have   constructed  four complete  evolutionary
calculation:

\begin{itemize}
\item Case A: Diffusion, all RLOF operate (including flash-induced RLOF)
\item  Case B:  Diffusion, without flash-induced RLOF
\item Case C: No diffusion, all RLOF operate
\item Case D: No diffusion, without flash-induced RLOF
\end{itemize}

Regarding mass  transfer episodes we have  chosen to study  the case of
fully non conservative conditions, i.e., those in which all the matter
transferred from  the primary  star is lost  from the  system carrying
away all  its intrinsic angular  momentum. We do  so in order  get the
strongest possible RLOFs which, in turn, will produce the largest mass
transfer episodes. In  this sense, we shall get an  upper limit to the
effects of  RLOFs on  the whole evolution  of the star,  in particular
regarding the ages of very cool WDs.

The  reminder  of the  paper  is  organized  as follows.   In  Section
\ref{sec:numerical} we describe our code paying special attention
to  the  changes we  implemented  in  the  scheme for  computing  mass
transfer episodes. Then, in  Section \ref{sec:calcula} we describe the
evolutionary results for the  four cases considered here.  Finally, in
\ref{sec:discu}  we discuss  the implicances  of our  calculations and
summarize the main conclusions of this work.

\section{Numerical Methods} \label{sec:numerical}

In  the computations  presented below  we have  employed the  code for
computing  stellar  evolution in  close  binary  systems presented  in
Benvenuto  \& De  Vito  (2003).   Now we  have  incorporated the  full
physical ingredients  with the aim of getting  state - of -  the - art
evolutionary results.  In particular,  we have included a complete set
of nuclear reactions to  describe hydrogen and helium burning together
with diffusion processes. For  more details on the considered physics,
see, e.g., Althaus, et al. (2001a).

Regarding  the  outer boundary  condition,  we  have incorporated  the
formula  given by  Ritter  (1988)  for  computing  the  mass  transfer  rate
$\dot{M}$:

\begin{equation}
\dot{M}= -\dot{M_{0}} \exp{\bigg( - \frac{R_{L}-R}{H_{P}} \bigg)};
\label{eq:ritter} \end{equation}

\noindent where $\dot{M_{0}}$ is the MTR for a star that exactly fills
the Roche  lobe (see Ritter's paper for its definition),
$R_{L}$  is the equivalent  radius of the  Roche lobe
(see below), $R$ is the stellar radius and $H_{P}$ is the photospheric
pressure scale height. We have considered that a mass transfer episode
is underway when $R \geq R_{L} - \xi H_{P}$ with $\xi=16$. In this way
the  star begins  (ends)  to transfer mass in  a very  natural and  smooth
way. This has  been completely adequate for the  purpose of carrying out
the calculations presented below.

\section{Evolutionary Calculations} \label{sec:calcula}

\begin{figure*} \epsfysize=600pt \epsfbox{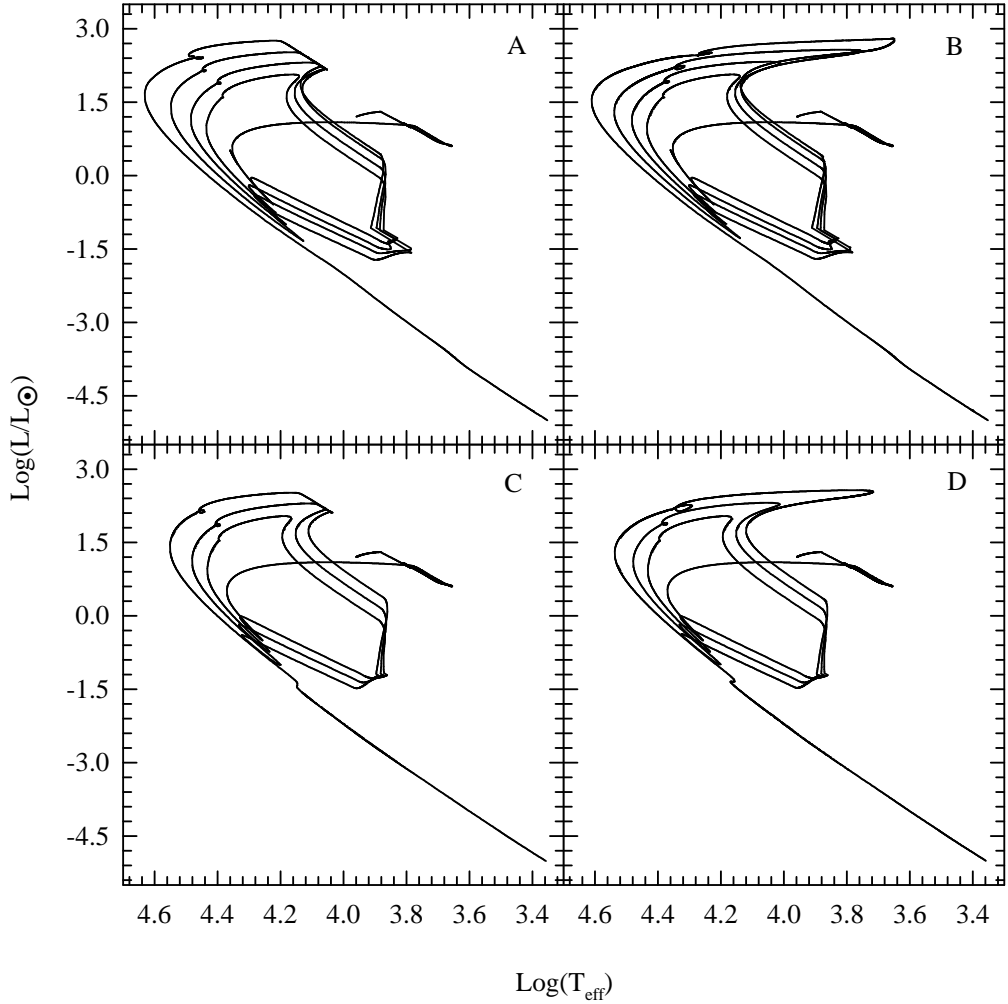} \caption{The 
evolutionary tracks  for  the primary  component of  the binary  system
initially  composed by  a normal  main  sequence, solar composition star 
of $M=  2.0$~\msun together with  a neutron  star of 1.4~\msun,  with  a 
period  of  $P=  1.0$ day  for  the  four  cases considered in this
paper. Each panel  is labeled as in the main text: {\bf  A}: diffusion,
all  RLOFs operate;  {\bf B}:  diffusion, no flash~-~induced RLOF
operates;  {\bf C}: no diffusion,  all RLOFs operate;  {\bf D}: no
diffusion, no flash~-~induced RLOF operates.  Notice that models  with
diffusion suffer from four hydrogen  thermonuclear flashes, while models
without diffusion undergo  only three.  This is independent  of
considering or ignoring the flash - induced mass transfer episodes. Here,
for the sake of clarity  we decided not to indicate the points given in
Tables~\ref{table:casoA}~-~\ref{table:casoC} }\label{fig:HRs}
\end{figure*}

\begin{figure*} \epsfysize=550pt \epsfbox{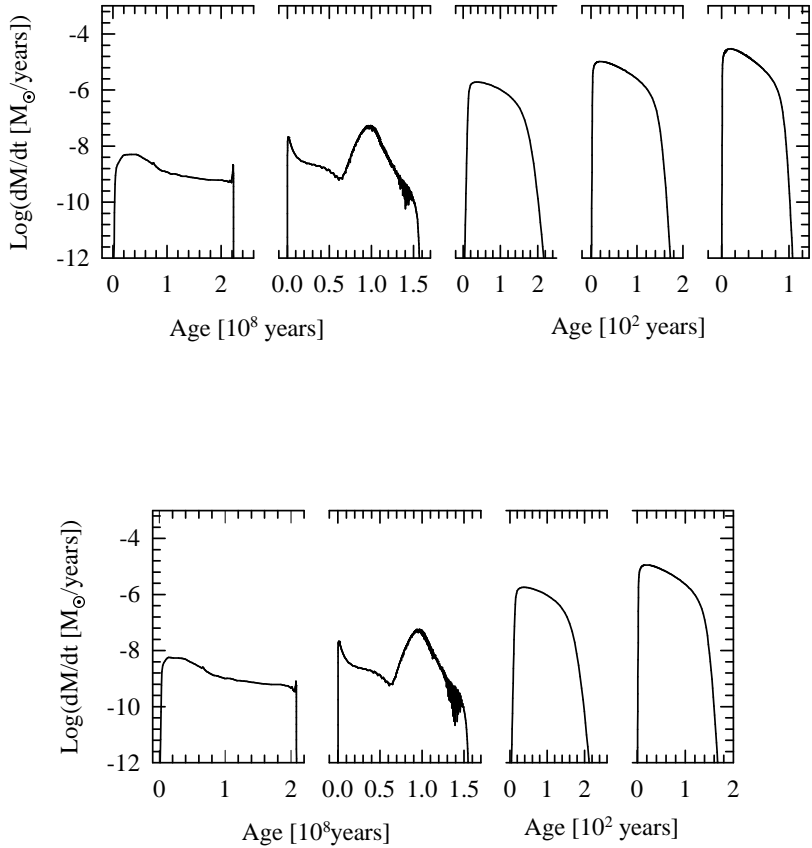} \caption{The mass 
transfer rates from the  primary star corresponding to the  evolutionary
tracks shown in Fig.~\ref{fig:HRs}  in which mass transfer is considered 
at each RLOF. Upper panel  corresponds to case {\bf A}  (with diffusion) 
while lower panel  depicts the  results for case  {\bf  C}  (without  
diffusion).   From  left  to  right  panels correspond to  each RLOF
episode. For  clarity, in each  panel we have counted time  since the
beginning  of each mass transfer  episode. For the absolute value of time
we simply have to add the time of the onset of  RLOF  given   in 
Tables~\ref{table:casoA}  and  \ref{table:casoC} respectively. Notice 
that, in  both cases, the  initial  RLOF is  very prolongated.   The first
happens  when the star is  still burning hydrogen at its  core. At core
exhaustion mass  transfer ends, and when the  star swells  due to  the
outwards  motion of  the  hydrogen shell burning, there  occurs a  new
RLOF. We called these two RLOFs as the {\it initial} RLOF. In  lower and
upper  panels these events  are almost  the same  due to the fact that
the  effects of  diffusion  are barely noticeable at these early stages.  
The subsequent episodes are due to thermonuclear  flashes. Notice that 
their duration  is six  orders of magnitude shorter. \label{fig:mlossA} }
\end{figure*}

\begin{figure*}
\epsfysize=550pt \epsfbox{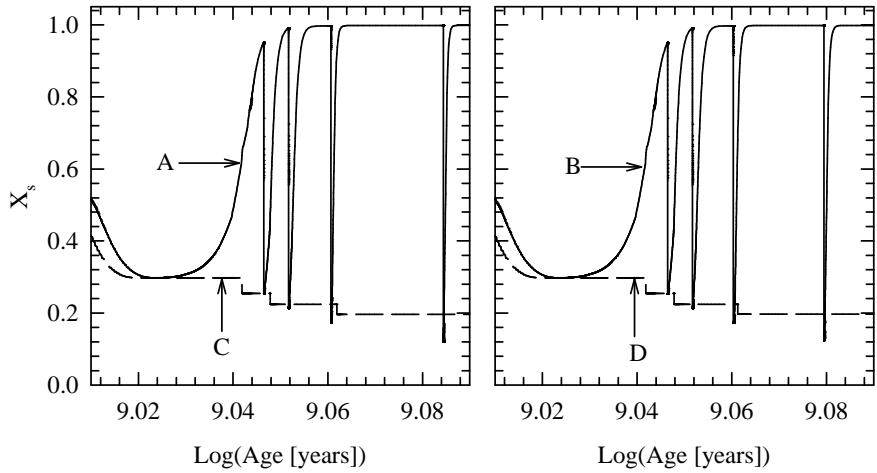}
\caption{The hydrogen abundance in the outermost layers of the primary
star.  Here we considered the hydrogen abundance at the first point in
the grid corresponding to  $1-M_r/M\approx~10^{-8}$. In the left panel
we depict  the results for models  in which RLOFs  are allowed whereas
right panel depicts the case  for models in which mass transfer driven
by thermonuclearly induced RLOFs  is neglected.  Each curve is labeled
as  in  Fig.~\ref{fig:HRs}. Notice  the  enormous  differences in  the
behaviour   of  outer   layers   in  the   cases   with  and   without
diffusion.  However, abundances  are barely  affected by  allowance of
mass transfer, as it is clearly noticeable in view of the similarities
of the plots in each panel.  In the sequences corresponding to case A,
due  to  mixing, mass  transfered  during  each  RLOF has  a  chemical
composition corresponding  to a minimum in hydrogen  abundance. Such a
composition is  very similar to that  of the plasma  transfered in the
models  of  Case  C  without  diffusion. See  main  text  for  further
details. \label{fig:abunda}}
\end{figure*}

In order  to present the  numerical calculations we shall  describe in
detail the sequence for which all physical ingredients were considered
(Case A) in which we allowed  diffusion and mass transfer in each RLOF to
operate.  The evolutionary track in the HR diagram corresponding to case
A  is shown in  Fig.~\ref{fig:HRs}. The 2~\msun object  begins to evolve
and the the first RLOF occurs (point 1   in
Table~\ref{table:casoA})  when it  is  still burning  hydrogen in  the
center.   Thus,  we are dealing with Class A mass  transfer, as defined 
by Kippenhahn \& Weigert (1967).  At that moment the hydrogen central 
abundance is $X_{H}=0.214578$. From there on, as consequence of the
orbital evolution of the binary, the primary star   undergoes  a   huge 
mass   loss  (see   the  first   panel  of Fig.~\ref{fig:mlossA})  which  
continues  up  to  the  moment (point  2  in Table~\ref{table:casoA})  
at   which   central  hydrogen   exhaustion occurs. The star contracts and
mass transfer is stopped; at that moment the  mass of  the  primary is 
of  1.59252 \msun.   Little latter,  as consequence of  the formation  of
a shell  hydrogen burning  zone, the star inflates and mass transfer 
starts again (point 3), and stands on a long period at which the star 
losses almost $90$ \% of the initial mass, ending with a mass of
0.22007~\msun. Hereafter we shall consider these {\it two} RLOF episodes
as an {\it initial} RLOF in order to differentiate it from the other
flash~-~induced RLOFs.  During  the initial mass  transfer   episode,  the
hydrogen  content  of  the outermost  layers  dropped up  to  a  minimum 
value of  $\approx~0.3$ because  mass   transfer  dredges  up  layers  
that  were  previously undergoing        appreciable        nuclear      
burning        (see Fig.~\ref{fig:abunda}). This increase in  the mean
molecular weight of the  plasma  present in  the  outer layers  of  the 
star favours  the contraction of the primary star. At point 4 the star
detaches from the Roche lobe, and since then on  the star evolves
bluewards very fast up to approximately  the moment at which  reaches a
local  maximum in the effective  temperature.   After  such maximum 
effective  temperature, evolution  appreciably  slows down  allowing 
diffusion  to have  time enough to sensibly evolve  the hydrogen profile. 
Here, hydrogen tends to  float simply  because it  is the  lighter
element  present  in the plasma.    This   is   clearly   shown   as  a  
steep   increase   in Fig.~\ref{fig:abunda}   where    we   show   the   
surface   hydrogen abundance\footnote{As a matter  of facts this is the 
abundance of the first  point   of  the  discrete   mesh  of  the 
model,   located  at $\log{1-M_r/M} \approx -8$.}  of the model as a
function of time.

\begin{centering}
\begin{table*}
\caption{\label{table:casoA}  Selected stages  of the  evolution  of a
system  composed by  a  1.4~\msun  neutron star  and  a normal,  solar
composition star of 2~\msun in orbit  with an initial period of 1 day.
Here we have  considered diffusion and mass transfer  during each RLOF
episode (Case  A).  Points labeled with odd  (even) numbers correspond
to   the   beginning   (end)   of   a   mass   transfer   episode   in
Fig.~\ref{fig:HRs}.   The last  point corresponds  to the  end  of the
computation.}
\begin{tabular}{cccccccccc} 
\hline
Point & $Log(L/L_{\odot})$ & $Log(L_{nuc}/L_{\odot})$ & $Log(T_{eff})$ & $X_{s}$  & $Age\; [Myr]$   & $M_{*}/M_{\odot}$ & $P\; [d] $ & $M_{H}/M_{*}$\\
\hline
\hline
    0 & 1.291596  & 1.291596  & 3.999353 & 0.700000 &     0.000000 &  2.00000 &  1.000 & 0.700000\\
    1 & 1.306425  & 1.325886  & 3.883615 & 0.719213 &   675.907133 &  2.00000 &  0.990 & 0.631826\\
    2 & 1.069236  & 1.085928  & 3.808278 & 0.700722 &   898.696085 &  1.59252 &  1.273 & 0.593744\\
    3 & 1.052254  & 1.083202  & 3.803819 & 0.700938 &   899.305975 &  1.59252 &  1.273 & 0.593691\\
    4 & 1.007000  & 0.998083  & 3.774765 & 0.297839 &  1057.346100 &  0.22007 &  4.309 & 0.029015\\
    5 & 2.158183  & 1.557922  & 4.071799 & 0.213738 &  1126.419320 &  0.22007 &  4.309 & 0.005668\\
    6 & 2.310374  & 1.558437  & 4.115358 & 0.213738 &  1126.419557 &  0.21990 &  4.310 & 0.005490\\
    7 & 2.151000  & 1.900315  & 4.068010 & 0.172838 &  1149.826078 &  0.21990 &  4.310 & 0.004322\\
    8 & 2.521391  & 1.900161  & 4.170841 & 0.172838 &  1149.826261 &  0.21919 &  4.313 & 0.003738\\
    9 & 2.210319  & 2.297033  & 4.080971 & 0.119486 &  1214.483908 &  0.21919 &  4.313 & 0.002783\\ 
   10 & 2.754083  & 2.360036  & 4.233749 & 0.119486 &  1214.484019 &  0.21801 &  4.320 & 0.002141\\  
   11 &-5.001480  & $-\infty$ & 3.352189 & 0.998488 & 18995.499104 &  0.21801 &  4.316 & 0.001609\\
\hline
\end{tabular}
\end{table*}
\end{centering}

\begin{figure*} \epsfysize=550pt \epsfbox{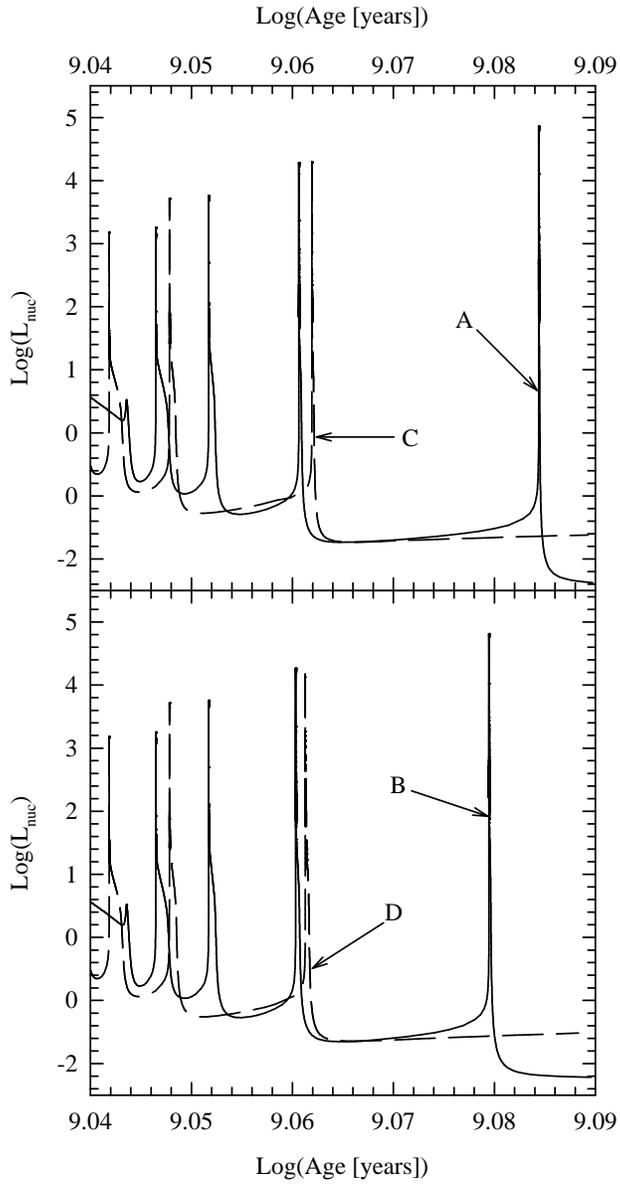} \caption{The
logarithm of the  nuclear luminosity (in solar units) vs. age relationship
for the four evolutionary sequences  studied in this paper.  Labels A, B,
C, D refer to the evolutionary tracks A to D in Fig.~\ref{fig:HRs}. Notice
that models  with (without) diffusion          undergo          four
(three)          thermonuclear flashes. \label{fig:flashes} }
\end{figure*}

Quite  in contrast  with the  behaviour of  surface abundance,  at the
bottom  of the  hydrogen  envelope, hydrogen  tend  to sink  now as  a
consequence  of large abundance  gradients.  The  net effect  is that,
while the  outer layers get  richer in hydrogen, diffusion  is fueling
hot layers.  Then, when the hydrogen rich layers become degenerate and
conduction eases the energy flux outwards, hydrogen becomes hot enough
to ignite.   Now, in degenerate conditions, ignition  is unstable (See
Fig.~\ref{fig:flashes}). Consequently,  evolution  suddenly  accelerates  and  there
occurs a  hydrogen thermonuclear  flash.  Such a  flash is  not strong
enough to  inflate the  star to force  a new RLOF.   Regarding surface
abundances,  we should  remark  that little  time after  thermonuclear
flash occurs, the star develops a deep outer convective zone embracing
from  very hydrogen  rich layers  up to  others in  which  hydrogen is
almost   a  trace.   As   consequence,  hydrogen   abundance  suddenly
drops\footnote{Notice  that this is  consequence of  the fact  that we
have assumed  instantaneous mixing throughout convective  zones.} down to a
value similar to  the one the star  had at the end of  the initial RLOF
(see   Fig.~\ref{fig:abunda}  and   Table   ~\ref{table:casoA}).   The
referred   mixing   is    noticeable   in   the   evolutionary   track
(Fig.~\ref{fig:HRs}) as a sudden change  of slope after the minimum in
the effective  temperature and luminosity of the  star.  After mixing,
the  outer  layers of  the  star continue swelling  up  to get  near
producing a RLOF, but begin  to contract before.  After maximum radius
is reached,  the star undergoes a  fast contraction up  to its maximum
effective temperature, and thereon again timescales become long enough
for  diffusion to  operate. The  star again  floats hydrogen  at outer
layers and  fuel others that  will make the  star to undergo  a second
thermonuclear flash. Now the star has a higher degree of degeneracy in
the critical layers, making the flash to be stronger than the previous
one.  From there  on the star undergoes much the  same evolution as in
the  previous flash, but  now the  star inflates  enough to  undergo a
third  RLOF event.  The  conditions at  the onset  of this  third RLOF
correspond to point 5 in Table ~\ref{table:casoA}.

Obviously, this third RLOF is deeply different to the initial one. Now the
envelope is very dilute and so, a tiny amount of stellar mass occupy a
large portion  of its radius.  Consequently, very little mass is
transferred from the primary in contrast with initial RLOF in which  the
primary transferred about 90\% of it initial mass.  The MTR during this
third RLOF is depicted in the third panel of  Fig.~\ref{fig:mlossA}. 
Remarkably,  the mass lost  from the primary star during  this third RLOF
has a  low hydrogen abundance due to the previous mixing.

\begin{figure*}
\epsfysize=550pt \epsfbox{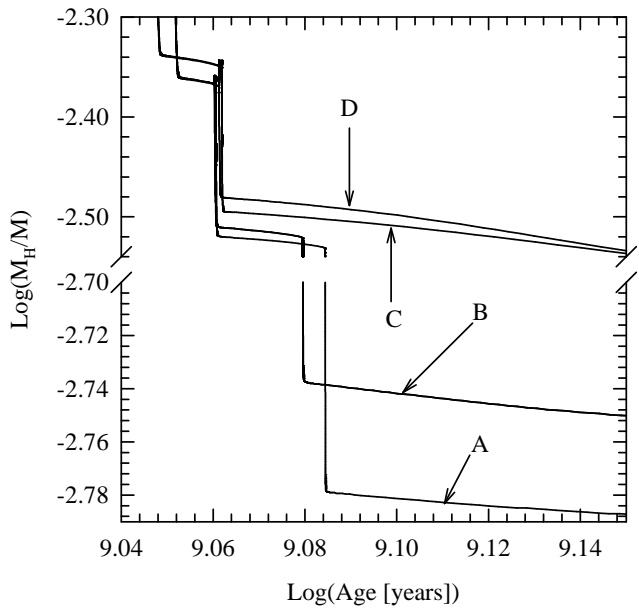}
\caption{The  logarithm of the  hydrogen mass  faction present  in the
star vs. time for the four  cases of evolution presented in this paper
during  the last  thermonuclear  flashes. Labels A, B, C, D refer to the evolutionary tracks A to D in Fig.~\ref{fig:HRs}.
Notice  that after thermonuclear  flashes, models
with  diffusion have a  lower hydrogen  fraction. Also,  as it  may be
expected, the hydrogen mass fraction is lower for models in which RLOFs
operate  at these  stages. Models  without diffusion  end  their flash
episodes with  a higher hydrogen  content which is  subsequently burnt
out during the final cooling  track. Notice the change in the vertical
scale  of  this figure  above  and below  the  break  in the  vertical
axis. See main text for further details. \label{fig:Hfinal} }
\end{figure*}

After  few thousands of  years transferring  mass, the  star contracts
again repeating essentially the same evolution the star followed after the
first thermonuclear flash.  However, remarkably, the star has lost
hydrogen,  due to nuclear  burning as  well as  to mass  transfer (see
Fig.~\ref{fig:Hfinal}).  However,  the star still  has an amount of
hydrogen high enough to  force the star to undergo another flash.  Now the
flash is rather more violent than the previous one producing another RLOF.

\begin{figure*}
\epsfysize=550pt \epsfbox{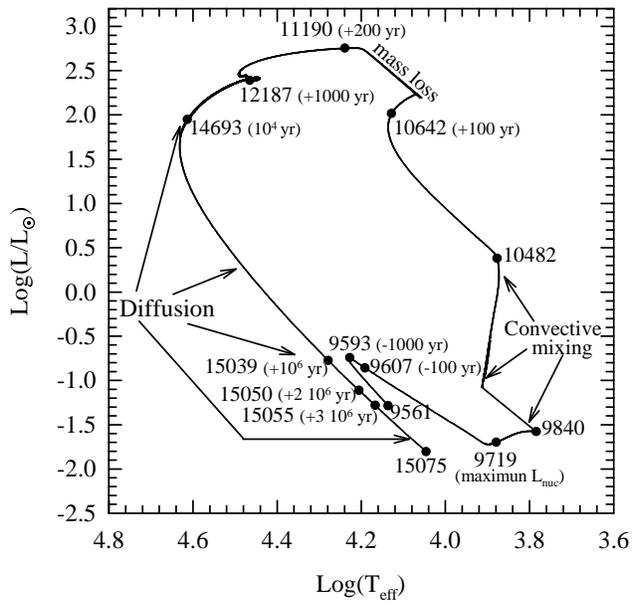}
\caption{The last loop in the HR diagram corresponding to case A
evolution.  Solid dots  correspond to  selected stages, for some of which the
hydrogen profiles are presented below  and are labeled with the number
of model  in the computed sequence.  Ages are counted  with respect to
the maximum  in the nuclear energy  release.  We also  remark the most
important physical  effects in each  portion of the track.  Notice the
sudden changes in the shape of  the track due to convective mixing and
mass loss.  \label{fig:loop} }
\end{figure*}

\begin{figure*}
\epsfysize=550pt \epsfbox{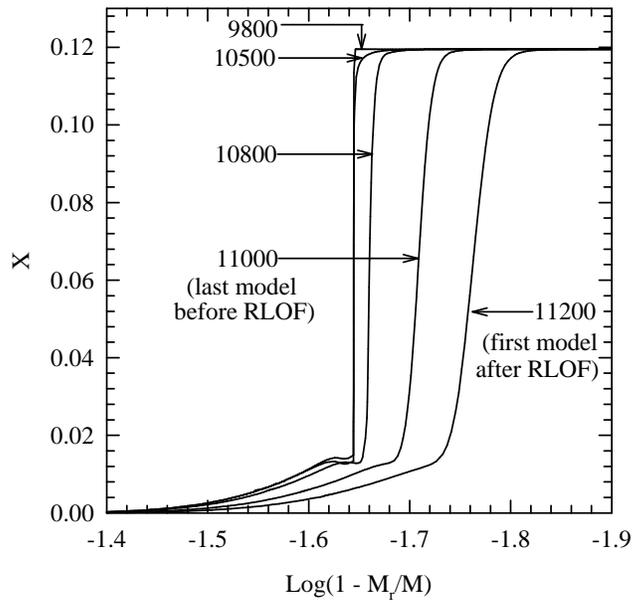}
\caption{The  hydrogen profile  for some  of the  models  indicated in
Fig.~\ref{fig:loop} corresponding to stages previous to and just after
the last  RLOF. Up to model  labeled 11000 the outwards  motion of the
profile is due to nuclear burning.  From this model to the one labeled
11200 the  motion is almost entirely  due to the  mass transfer during
the RLOF. \label{fig:perfiles_premloss} }
\end{figure*}

\begin{figure*}
\epsfysize=550pt \epsfbox{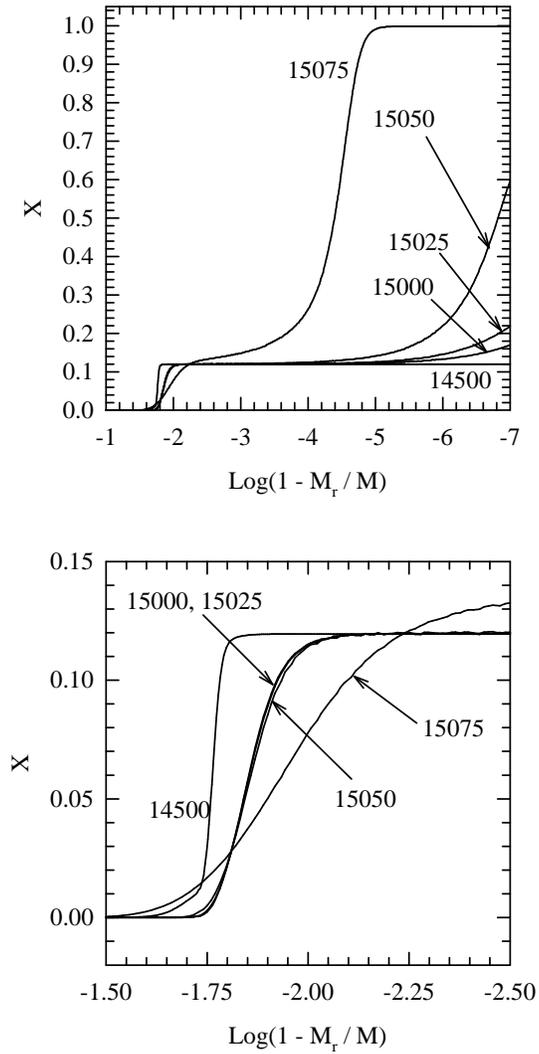}
\caption{The  hydrogen profile  for some  of the  models  indicated in
Fig.~\ref{fig:loop} corresponding to stages after the last RLOF. Upper
panel  represents  the whole  hydrogen  profile  while  the lower  one
highlights the bottom of the profile. The evolution of the outer parts
of the profile are fully dominated by diffusion. In the bottom there is
an interplay between nuclear burning and diffusion.  Initially nuclear
reactions dominate (models 14500 to  15050), but at later times (model
15075) chemical   diffusion  drives   a  large   amount   of  hydrogen
inwards. \label{fig:perfiles_postmloss} }
\end{figure*}

In order to gain clarity, we have chosen to discuss in detail the loop
due to  the last  thermonuclear flash.  In Fig.~\ref{fig:loop} we
show the excursion  of the star in the HR  diagram indicating the main
physical  agents acting  in  the star  together  with some  particular
models (solid  dots).  The  hydrogen profiles corresponding  to models
before and just at the end of  (after) the RLOF  are shown  in  Fig.~\ref{fig:perfiles_premloss}
(Fig.~\ref{fig:perfiles_postmloss}).  Some relevant characteristics of
these models  are presented in Table~\ref{table:difu}.

The  hydrogen profile  for some  of the  models  indicated in
Fig.~\ref{fig:perfiles_premloss} corresponding to stages previous to and just after
the last  RLOF. Up to model  labeled 11000 (the model
number  in the sequence)  the outwards  motion of the
profile is due to the nuclear burning during the flash. Since model 11000 on (not
shown in the figure) the profile moves outwards as consequence of the mass
transfer episode which ends in the model 11200. Notice that the loss of 
hydrogen is somewhat tiny (points 9 and 10 in Table~\ref{table:casoA}).

In
Fig.~\ref{fig:perfiles_postmloss}  curve labeled  as 14500  corresponds to  stages somewhat  after RLOF.
Curves labeled as 15000 -  15050 are displaced outwards due to nuclear
burning  while  curve  15075  corresponds  to a  profile  modified  by
diffusion. Notice  the tail of  the hydrogen profile  gets appreciably
deeper  approximately when  the outermost  layers become  saturated by
hydrogen.

\begin{centering}
\begin{table*}
\caption{\label{table:difu}  Selected stages of  the evolution  of the
primary  star of  the system  corresponding to Case  A. We  have  selected relevant
models in  a loop  shown at Fig.~\ref{fig:loop}.  Model stand  for the
number of models in the evolutionary calculation.}
\begin{tabular}{rcccccccc}
\hline   Model   &   $Log(L/L_\odot)$   &   $Log(L_{nuc}/L_\odot)$   & $Log(T_{eff})$ &  $X_{s}$ & Age [Myr] &  $M_{H}/M_{*}$\\ 
\hline \hline
9561  & -1.285417 & -1.464898 & 4.136383 & 0.997875 & 1206.136861 & 0.002962\\  
9593  & -0.743989 &  1.319834 & 4.227583 & 0.997875 & 1214.482831 & 0.002921\\  
9607  & -0.859529 &  2.616561 & 4.191011 & 0.997875 & 1214.483706 & 0.002915\\  
9719  & -1.700589 &  4.860120 & 3.878964 & 0.997875 & 1214.483802 & 0.002820\\  
9840  & -1.577785 &  4.118219 & 3.784233 & 0.997875 & 1214.483809 & 0.002790\\  
10482 &  0.378350 &  2.338424 & 3.877158 & 0.119486 & 1214.483895 & 0.002783\\
10642 &  2.014473 &  2.313509 & 4.127957 & 0.119486 & 1214.483902 & 0.002783\\  
10714 &  2.210319 &  2.297033 & 4.080971 & 0.119486 & 1214.483908 & 0.002783\\  
11190 &  2.754083 &  2.360036 & 4.233749 & 0.119486 & 1214.484019 & 0.002141\\  
12187 &  2.390152 &  2.739479 & 4.464822 & 0.119486 & 1214.484802 & 0.002120\\  
14693 &  1.948789 &  1.939478 & 4.613450 & 0.120118 & 1214.493795 & 0.002062\\  
15039 & -0.774868 & -1.657584 & 4.278570 & 0.395631 & 1215.485203 & 0.001664\\
15050 & -1.112451 & -2.015089 & 4.205475 & 0.720579 & 1216.420781 & 0.001664\\  
15055 & -1.281705 & -2.151893 & 4.167198 & 0.921884 & 1217.429788 & 0.001663\\ 
15075 & -1.805696 & -2.445395 & 4.045660 & 0.998488 & 1267.968504 & 0.001653\\ 
\hline
\end{tabular}
\end{table*}
\end{centering}

\begin{figure*} \epsfysize=550pt \epsfbox{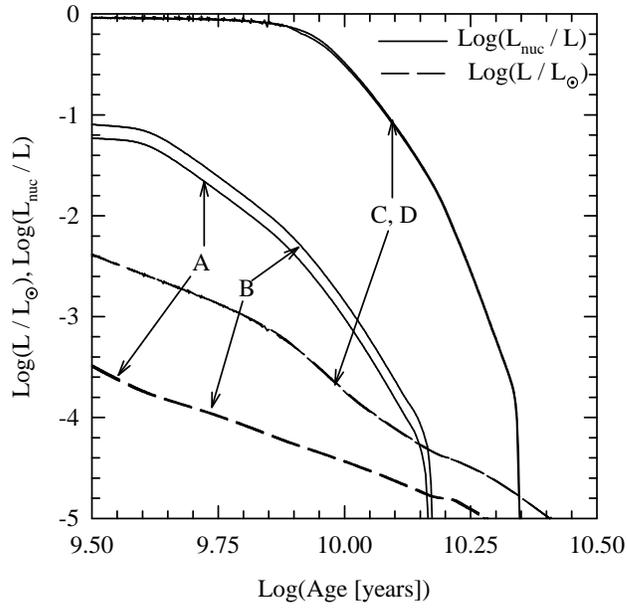} \caption{The 
logarithm of  photon luminosity  and  nuclear luminosity fraction
released  during the last  stages of evolution of  the models considered
in this paper.  Labels A, B, C, D refer to the evolutionary tracks A to D
in Fig.~\ref{fig:HRs}. Models  without  diffusion have  a  much  larger 
nuclear activity  at advanced  evolutionary stages, which  slows down  the
evolution  in an appreciable  way. Remarkably, the  ages of  these
objects  are largely determined  by  the  allowance  of diffusion 
whereas  considering  or neglecting thermonuclearly  induced RLOFs  has a
negligible  effect on stellar ages. \label{fig:lumi} } \end{figure*}

From  our calculations  we find  that only  after {\it  four} hydrogen
thermonuclear flashes the  star is able to enter  on the final cooling
track of  a helium WD (see  Fig.~\ref{fig:flashes}).  The evolutionary
timescale of  the model  is presented in  Fig.~\ref{fig:lumi}.  Notice
that the nuclear  energy release at such advanced  stages of evolution
is a minor  contribution to the total energy balance  of the star.  As
consequence the  star gets heat  from its relic thermal  content which
forces a fast evolution,  reaching very low luminosities in timescales
comparable to the age of the Universe. We stopped the calculation when
the object reached \ll_lsun=~-5, at that moment the star had an age of
about 19 Gyr.

%
%

Now, let  us discuss the results  corresponding to Case C  in which we
allowed  all  RLOF  to  drive  mass transfer  but  we  have  neglected
diffusion (see  Table~\ref{table:casoC}).  The evolutionary  track for
this case is shown in Fig.~\ref{fig:HRs}, panel C. Here, the evolution
previous  to the  end  of the  initial  RLOF is  very  similar to  that
corresponding  to  the  Case  A.   In other  words,  diffusion  has  a
negligible  effect on  these evolutionary  stages.  Perhaps,  the main
difference is the increase  in hydrogen surface abundances previous to
the   initial   RLOF   found   in   Case   A   is   absent   here   (see
Table~\ref{table:casoA}).  However,  differences in the  evolution are
quite significantly after the end of initial RLOF.

As here, by assumption, the  physical agents able to modify abundances
are only  nuclear reactions  and convection, obviously,  the outermost
layers of  the hydrogen - rich  layers are not  enriched by hydrogen
and simultaneously no fueling  occurs at the bottom. Consequently, the
evolution is  very different.  Notably, the star  undergoes only three
thermonuclear flashes and outer layers have rather constant abundances
in  spite that  they also  develop outer  convection zones  after each
flash. As  in the previous case  we computed the evolution  up to when
the object reached \ll_lsun=~-5 with  an age of about 25.65 Gyr. Thus,
evolution is markedly  slower than in Case A. This is  due to the fact
that here thermonuclear flashes  are weaker, burning less hydrogen. As
consequence,  the star  is able  to undergo  appreciable thermonuclear
energy release during the final cooling  track as a helium WD. In this
regard, notice that nuclear burning remains the dominant energy source
of the star up to ages of 10 Gyr (see Fig.~\ref{fig:lumi}).

\begin{centering}
\begin{table*}
\caption{\label{table:casoC} Selected stages  of the evolution  of a system composed  by a
1.4~\msun neutron star and a normal, solar composition star of 2~\msun
in orbit with  a period of 1 day.  Here mass  transfer is allowed to occur during
each RLOF  episode but diffusion  has been neglected (Case  C). Points
labeled with odd (even) numbers correspond to the beginning (end) of
a  mass  transfer  episode.  The  last  point
corresponds to the end of the computation.}
\begin{tabular}{cccccccccc} 
\hline
Model   & $Log(L/L_{\odot})$ & $Log(L_{nuc}/L_{\odot})$ & $Log(T_{eff})$ & $X_{s}$  & $Age\; [Myr]$   & $M_{*}/M_{\odot}$ & $P\; [d] $ & $M_{H}/M_{*}$\\
\hline
\hline
   0  & 1.291596  & 1.291596 & 3.999353 & 0.700000 &      0.000000 &  2.00000 & 1.000 & 0.700000\\
   1  & 1.305931  & 1.325651 & 3.883688 & 0.700000 &    683.258119 &  2.00000 & 0.990 & 0.631100\\
   2  & 1.080234  & 1.081842 & 3.811255 & 0.699873 &    892.518745 &  1.60119 & 1.266 & 0.595018\\
   3  & 1.059051  & 1.092015 & 3.805732 & 0.699873 &    893.302911 &  1.60119 & 1.266 & 0.594950\\
   4  & 1.011273  & 0.998631 & 3.776142 & 0.297567 &   1049.362320 &  0.22033 & 4.307 & 0.028800\\
   5  & 2.147837  & 1.441249 & 4.068767 & 0.224255 &   1116.384319 &  0.22033 & 4.307 & 0.005821\\
   6  & 2.297723  & 1.446788 & 4.111879 & 0.224255 &   1116.384550 &  0.22018 & 4.308 & 0.005645\\
   7  & 2.099698  & 1.812834 & 4.054606 & 0.196021 &   1153.232534 &  0.22018 & 4.308 & 0.004501\\
   8  & 2.520010  & 1.808553 & 4.171450 & 0.196021 &   1153.232711 &  0.21943 & 4.312 & 0.003816\\
   9  &-5.011041  & $-\infty$ & 3.356698 & 0.196019 & 25653.632359 &  0.21943 & 4.308 & 0.001758\\
\hline
\end{tabular} 
\end{table*}
\end{centering}

The  sequence of  models  corresponding to  Case  B (Case  C) is  very
similar  to that  corresponding to  Case A  (Case D)  and will  not be
discussed in detail.   The obvious major difference is  related to the
size the star is able to reach just after thermonuclear flashes. As in
Case B (case D) it is assumed  that there is no limitation imposed by the
size of the Roche lobe,  after thermonuclear flashes, the star reaches
effective temperatures far lower than those allowed in the case of the
occurrence of  RLOFs. Quite noticeably, the  evolutionary timescale of
the final WD cooling track is largely independent of the occurrence of
any thermonuclearly flash - induced RLOF (see Fig.~\ref{fig:lumi}).

Another  interesting  difference  arises regarding  the  characteristic
timescale of evolution of the models  from the red part of the diagram
to the maximum effective temperature conditions. We have found that in
such  stages, evolution  of  models for  which thermonuclearly  induced
RLOFs are  allowed suffer from  a much faster evolution  regardless of
the allowance or not of diffusion. To  be specific, from minimum  to maximum
effective  temperature, Case  B models  take $\approx  10^{6}$~y while
models  with  RLOF  spend  only  $\approx 10^{4}$~y  from  Roche  lobe
detachment   to  maximum   effective  temperature.   This,  obviously
indicates that  it should  be more difficult  to find objects  at such
conditions than predicted in  models without thermonuclearly - induced
RLOFs. Also, notice in  Fig.~\ref{fig:HRs} that the subflashes (little
loops in the  evolutionary tracks occurring when the  star is evolving
bluewards) happen at different effective temperatures depending on the
allowance of RLOFs. In fact, in models with (without) RLOFs subflashes
occur  at  higher   (lower)  effective  temperatures  for  consecutive
thermonuclear flashes. This  is so irrespective of allowance  or not of
diffusion.
 
\section{Discussion and Conclusions} \label{sec:discu}

In  this  work we  have  computed the  evolution  of  a binary  system
composed by a neutron star  with a ``canonical mass'' of 1.4~\msun and
a normal, population I main sequence star of 2~\msun in orbit with a 1
day period.   We have performed the calculations  employing an updated
version of the code presented in  Benvenuto \& De Vito (2003) in which
we have included the  main standard physical ingredients together with
diffusion  processes.   Also a  proper  outer  boundary condition  was
incorporated following Ritter (1988) (see \S \ref{sec:numerical}).

In  order to  explore  the role  of  mass transfer  episodes from  the
primary star and its interplay  with diffusion we have considered four
situations:  diffusion,  all RLOF operate (Case  A); diffusion, no
flash~-~induced RLOF operates (Case  B); no diffusion, all RLOF  operate
(Case C); and no  diffusion, no flash~-~induced RLOF operates   (Case 
D).  See  introduction (\S~\ref{sec:intro}) of further details.

To our  notice, these calculations represent the  first detailed study
of binary  evolution considering diffusion.  In this sense,  this work
represents  a  natural  generalization  of the  results  presented  by
Althaus, Serenelli \& Benvenuto  (2001a) in which binary evolution
processes was  mimicked by  forcing a 1~\msun  star on the  red giant
branch  to undergo  an appropriate  mass  loss rate.   Now the  proper
inclusion of the specific processes that govern binary evolution offer
us a more  physically sound description of the  formation of low mass,
helium white dwarfs (WDs). In  particular, now we have the possibility
of  connecting  stellar  structure  and  evolution  with  the  orbital
parameters  of  the systems  allowing  for  a  deeper comparison  with
observations.

From the results presented in  the previous sections, it is clear that
diffusion  is  far more  important  in  determining  the timescale  of
evolution of the stars than  mass transfer episodes during flash~-~induced
RLOFs. This is so especially  when the object reaches the  final cooling
track. We found  that timescales are  almost insensitive  to the
occurrence of flash~-~induced RLOF  episodes (see Fig.~\ref{fig:lumi}).
This constitutes the  main result of the present work.

\section{acknowledgments}

We thank our referee, Prof. Philipp Podsiadlowski for comments and suggestions
that allowed to improve the clarity of the original version. OGB is supported
by FONDAP Center for Astrophysics 15010003.



\bsp

\label{lastpage}

\end{document}